\documentclass[11pt]{article}
\usepackage{amssymb,latexsym,amsmath}
\usepackage[dvips]{graphicx}
\usepackage{cite}
\newtheorem{theo}{Theorem}

\newtheorem{prop}[theo]{Proposition}
\def\nn{\nonumber}
\def\deg{\mathop{\rm deg}\nolimits}

\def\qdots{\mathinner{\mkern1mu\raise1pt\vbox{\kern7pt\hbox{.}}\mkern2mu
 \raise4pt\hbox{.}\mkern2mu\raise7pt\hbox{.}\mkern1mu}}
\def\Z{{\mathbb Z}}

\def\R{{\mathbb R}}

\def\gl{\mathfrak{gl}}
\def\ssl{\mathfrak{sl}}

\def\lb{[\![}
\def\rb{]\!]}
\newcommand{\q}{\hat q}
\newcommand{\p}{\hat p}

\def\mybox{\hfill$\Box$}
\renewcommand{\theequation}{\arabic{section}.\arabic{equation}}

\begin{document}
\begin{center}
{\Large \bf
On the eigenvalue problem for arbitrary odd elements of  \\[2mm]
the Lie superalgebra $\gl(1|n)$ and applications}\\[5mm]
{\bf S.~Lievens\footnote{E-mail: Stijn.Lievens@UGent.be}, }
{\bf N.I.~Stoilova}\footnote{E-mail: Neli.Stoilova@UGent.be; Permanent address:
Institute for Nuclear Research and Nuclear Energy, Boul.\ Tsarigradsko Chaussee 72,
1784 Sofia, Bulgaria} {\bf and J.\ Van der Jeugt}\footnote{E-mail:
Joris.VanderJeugt@UGent.be}\\[1mm]
Department of Applied Mathematics and Computer Science,
Ghent University,\\
Krijgslaan 281-S9, B-9000 Gent, Belgium.
\end{center}


\begin{abstract}
In a Wigner quantum mechanical model, with a solution in terms of the Lie superalgebra $\gl(1|n)$,
one is faced with determining the eigenvalues and eigenvectors for an arbitrary
self-adjoint odd element of $\gl(1|n)$ in any unitary irreducible representation $W$.
We show that the eigenvalue problem can be solved by the decomposition of $W$ with
respect to the branching $\gl(1|n) \rightarrow \gl(1|1)\oplus \gl(n-1)$. 
The eigenvector problem is much harder, since the Gel'fand-Zetlin basis of $W$
is involved, and the explicit actions of $\gl(1|n)$ generators on this
basis are fairly complicated. 
Using properties of the Gel'fand-Zetlin basis, we manage to present a solution for this 
problem as well.
Our solution is illustrated for two special classes of unitary $\gl(1|n)$ representations:
the so-called Fock representations and the ladder representations.
\end{abstract}

%

\setcounter{equation}{0}
\section{Introduction}
\label{sec:intro}

Recently, the Wigner quantum approach of a
quantum mechanical model consisting of a linear chain of $n$ identical harmonic 
oscillators coupled by some nearest neighbour interaction was considered~\cite{LSV}.
In the standard approach, where the canonical commutation relations
between position and momentum operators are required, a solution of the
system is well known~\cite{Cohen}. 
In~\cite{LSV} it was shown that these requirements can be relaxed and the
problem was treated as a Wigner quantum system.
As a consequence, the system allows besides the canonical solution also
other types of solutions. In particular, it was shown that the (finite-dimensional) unitary
irreducible representations of the Lie superalgebra $\gl(1|n)$ give
rise to new solutions.

In order to study properties of these new solutions, one is faced with some 
computationally difficult problems in the representation theory of $\gl(1|n)$~\cite{Kac1,Kac2}.
More precisely, consider the standard basis of $\gl(1|n)$ consisting of
elements $e_{ij}$ ($0\leq i,j\leq n$), with $e_{0j}$ and $e_{j0}$ ($1\leq j\leq n$)
the odd elements of the Lie superalgebra, with bracket~(\ref{eij}), and with
star condition $e_{ij}^\dagger = e_{ji}$. 
The unitary representations $W=W([m]_{n+1})$ of $\gl(1|n)$ are well known~\cite{Gould}: 
they are labeled by
some $(n+1)$-tuple $[m]_{n+1}$ subject to certain conditions.
Even more: for such representations, a Gel'fand-Zetlin basis has been constructed
and the explicit action of the $\gl(1|n)$ generators on the basis vectors
of $W$ is also known~\cite{KSV}. 
Explicit actions of generators on a Gel'fand-Zetlin basis (GZ-basis) are usually quite involved,
and this is also the case for $\gl(1|n)$. In particular, the action of
the odd generators $e_{0j}$ and $e_{j0}$ on a GZ-basis vector is very
complicated, see~(\ref{e0j})-(\ref{ej0}).

The operators we intend to study are the position and momentum operators $\q_r$ and $\p_r$
($r=1,\ldots,n$) of the quantum system. These are self-adjoint operators, and in
the $\gl(1|n)$ solution of the problem considered in~\cite{LSV} as a Wigner quantum system their expression
is of the form
\begin{equation}
\sum_{j=1}^n \alpha_j\; e_{0j} + \sum_{j=1}^n \alpha_j^*\; e_{j0},
\label{arbitrary}
\end{equation}
for certain constants $\alpha_j$. This is an arbitrary self-adjoint odd element in $\gl(1|n)$.
For such elements, we want to determine the spectrum (eigenvalues) in any
unitary representation $W$. Furthermore, we wish to construct an explicit
set of orthonormal eigenvectors of~(\ref{arbitrary}) in terms of the GZ-basis
of $W$.
The eigenvalue problem turns out to be feasible, thanks to group theoretical methods. 
In fact, we show how it is related to the decomposition of $\gl(1|n)$ representations
into representations of the subalgebra $\gl(1|1)\oplus \gl(n-1)$.
The eigenvector problem is much harder, as one is faced with the sophisticated
action of the Lie superalgebra generators on the GZ-basis vectors. 
But also here, we manage to present a solution.

The structure of the paper is as follows.
In Section~\ref{sec:problem} we describe in more detail the origin of the problem.
We also recall the structure of the GZ-basis for $\gl(1|n)$ representations,
and the conditions for unitarity. 
In Section~\ref{sec:another} we convert the general eigenvalue problem to 
a simpler problem by switching to another set of odd generators for
the Lie superalgebra $\gl(1|n)$. 
In terms of the new set of generators, (\ref{arbitrary}) has a simple
expression: in fact it becomes an element of a $\gl(1|1)$ subalgebra
of $\gl(1|n)$. The branching $\gl(1|n) \rightarrow \gl(1|1)\oplus \gl(n-1)$,
studied in Section~\ref{sec:decomposition}, leads to an answer
of the eigenvalue problem. 
In the next section, we construct the essential relation that expresses
the highest weight vector of $W$ with respect to the new set of generators
in terms of the ``old'' GZ-basis vectors. 
Combining this with the known actions on such GZ-basis vectors yields
a solution for the eigenvector problem.
Then we illustrate our results for two important classes of unitary representations.
Section~\ref{sec:Wp} deals with Fock representations of $\gl(1|n)$. These
representations are quite simple, and have been considered in~\cite{Palev3}. 
Nevertheless, the eigenvector problem turned out to be difficult and
was left as an open problem in~\cite{LSV}. With the techniques developed
in this paper, a simple solution to the eigenvector problem is obtained.
Section~\ref{sec:Vp} deals with another relatively simple class of representations,
the ladder representations of $\gl(1|n)$. Also here, we illustrate how
our techniques lead to a complete solution of the eigenvalue and eigenvector problem.
We conclude the paper by some final remarks.

\setcounter{equation}{0}
\section{Description of the problem}
\label{sec:problem}

In several models~\cite{Audenaert,Brun,Eisert,Halliwell,Plenio} a quantum system consisting of 
a linear chain of $n$ identical harmonic oscillators
coupled by springs is used. The Hamiltonian of such a system is given by:
\begin{equation}
\hat{H}=\sum_{r=1}^{n} \Big( \frac{\hat{p}_r^2}{2\mu}
+ \frac{\mu\omega^2}{2} \hat{q}_r^2 + \frac{c\mu}{2}(\hat{q}_r-\hat{q}_{r+1})^2  \Big),
\label{Intro-H}
\end{equation}
where each oscillator has mass $\mu$ and frequency $\omega$, $\q_r$ and $\p_r$ stand
for the position and momentum operator for the $r$th oscillator (or rather,
$\q_r$ measures the displacement of the $r$th mass point with respect to its
equilibrium position), and $c>0$ is the coupling strength. Often, one assumes
periodic boundary conditions (also in this paper), i.e.
\begin{equation}
\q_{n+1}\equiv \q_1.
\label{qn+1}
\end{equation}
In the solution for such a system, one introduces finite Fourier transforms of the
(self-adjoint) operators $\q_r$ and $\p_r$ by
\begin{align}
\q_r & = \sum_{j=1}^n \sqrt{\frac{\hbar}{2\mu n\omega_j}} \left( e^{-2\pi i jr/n} a_j^+ +
e^{2\pi i jr/n} a_j^- \right), \\
\p_r & = \sum_{j=1}^n i\; \sqrt{\frac{\mu\omega_j\hbar}{2n}} \left( e^{-2\pi i jr/n} a_j^+ -
e^{2\pi i jr/n} a_j^- \right),
\end{align}
where $\omega_j$ are positive numbers with 
\begin{equation}
\omega_j^2=\omega^2 +2c-2c\cos(\frac{2\pi j}{n})=\omega^2+4c\sin^2(\frac{\pi j}{n}),
\label{omega}
\end{equation}
and $a_j^\pm$ are operators satisfying $(a_j^\pm)^\dagger = a_j^\mp$. In terms of these new
operators, the Hamiltonian reads
\begin{equation}
\hat{H}=\sum_{j=1}^{n} \frac{\hbar \omega_j}{2} (a_j^- a_j^+ + a_j^+ a_j^-). 
\label{Haa}
\end{equation}
If one assumes the canonical commutation relations for the operators $\q_r$ and $\p_r$,
then the operators $a_j^\pm$ satisfy the usual boson relations $[a_j^\pm,a_k^\pm]=0$, $[a_j^-,a_k^+]=\delta_{jk}$,
and the corresponding solutions are easy to describe. 
In~\cite{LSV} however, it was shown that one can relax the canonical commutation relations
for this system, and instead approach it as a Wigner quantum system~\cite{Palev86}-\cite{Palev1}, leading to other classes
of solutions besides the canonical ones. In this approach, the canonical commutation relations
are not required but replaced by the quantization relations following from the compatibility
between Hamilton's equations and the Heisenberg equations. Explicitly, these relations are~\cite{LSV}
\begin{equation}
\Bigl[ \sum_{j=1}^{n}  \omega_j (a_j^- a_j^+ + a_j^+ a_j^-) , a_k^\pm \Bigr]=
\pm 2 \omega_k a_k^\pm, \quad \ (k=1,2\ldots,n). 
\label{algrelations}
\end{equation}
These are triple relations involving anticommutators and commutators, and it was shown~\cite{LSV} that 
such relations have a solution in terms of generators of the Lie superalgebra $\gl(1|n)$~\cite{Kac1,Kac2}. 
More explicitly, let $\gl(1|n)$ be the Lie superalgebra  with standard basis
elements $e_{jk}$ ($j,k=0,1,\ldots,n$) where $e_{k0}$ and $e_{0k}$ ($k=1,\ldots,n$)
are odd elements and the remaining basis elements are even, with bracket
\begin{equation}
\lb e_{ij}, e_{kl} \rb = \delta_{jk} e_{il} - (-1)^{\deg(e_{ij})\deg(e_{kl})} \delta_{il}e_{kj},
\label{eij}
\end{equation}
and star condition $e_{ij}^\dagger=e_{ji}$. 
Then a solution of~(\ref{algrelations}) is provided by
\begin{equation}
a_j^- = \sqrt{\frac{2\beta_j}{\omega_j}}\;  e_{j0}, \quad
a_j^+ = \sqrt{\frac{2\beta_j}{\omega_j}}\;  e_{0j}, \quad(j=1,\ldots,n)
\label{solution}
\end{equation}
where
\begin{equation}
\beta_j = -\omega_j + \frac{1}{n-1}\sum_{k=1}^n \omega_k, \quad (j=1,\ldots,n).
\label{beta}
\end{equation}
All these numbers $\beta_j$ should be nonnegative, and in~\cite{LSV} we have shown that
this is possible provided the coupling constant $c$ is bounded by some critical value $c_0$
(which we shall assume to be the case here).
So, for this $\gl(1|n)$ solution, one finds
\begin{align}
\q_r & = \sqrt{\frac{\hbar}{\mu n}} \sum_{j=1}^n \left( \gamma_j\, e^{-2\pi i jr/n} e_{0j} +
\gamma_j\, e^{2\pi i jr/n} e_{j0} \right), 
\label{q-e}\\
\p_r & = i\; \sqrt{\frac{\mu\hbar}{n}}\sum_{j=1}^n  \left( \sqrt{\beta_j}\;e^{-2\pi i jr/n} e_{0j} -
\sqrt{\beta_j}\; e^{2\pi i jr/n} e_{j0} \right),
\label{p-e}
\end{align}
where we introduce yet another set of positive numbers
\begin{equation}
\gamma_j = \sqrt{\beta_j}/\omega_j \quad (j=1,\ldots,n) \hbox{ and }
\gamma = \gamma_1^2 + \cdots + \gamma_n^2.
\end{equation}
Equations~(\ref{q-e}) and~(\ref{p-e}) give a description of the ``physical operators'' $\q_r$
and $\p_r$ in terms of $\gl(1|n)$ generators. In order to study properties of such
operators (spectra or eigenvalues, eigenvectors), one should consider representations of
$\gl(1|n)$ for which the star condition $e_{ij}^\dagger=e_{ji}$ is satisfied.
These are the star representations or unitary representations $W([m]_{n+1})$ of $\gl(1|n)$, and they
are well known~\cite{Gould}. 

As (\ref{q-e}) and (\ref{p-e}) are the ``physical operators'' corresponding to 
position and momentum of the $r$th oscillator, we are interested in 
the following problems:
\begin{itemize}
\item[(a)] describe the eigenvalues of $\q_r$ and $\p_r$ in any unitary representation $W([m]_{n+1})$;
\item[(b)] construct the eigenvectors of $\q_r$ and $\p_r$ in $W([m]_{n+1})$.
\end{itemize}
At first sight, these problems might look easy.
However, a closer look at the explicit actions of the $\gl(1|n)$ generators $e_{0j}$ and $e_{j0}$
on the GZ-basis, see eqs.~(2.25) and (2.26) in~\cite{KSV} or~(\ref{e0j})-(\ref{ej0}) in
the Appendix, shows that these expressions are
extremely complicated. Since (\ref{q-e}) and (\ref{p-e}) are linear combinations of the 
generators $e_{0j}$ and $e_{j0}$, one could expect that the answer to the above questions
gives rise to unfeasible computations.

Nonetheless, we shall show that a group theoretical approach (using subalgebras, branching
rules, and a proper use of two inequivalent GZ-bases) leads to a solution for these two problems.

We end this section by describing the relevant representations, i.e.\ the unitary irreducible
representations $W([m]_{n+1})$ of $\gl(1|n)$~\cite{Gould}, and their GZ-basis~\cite{KSV}.
The finite-dimensional irreducible representations (simple modules) $W([m]_{n+1})$ of the Lie 
superalgebra $\gl(1|n)$ are in one-to-one correspondence with the set 
of all complex $(n+1)$-tuples~\cite{Kac1,Kac2}
\begin{equation}
[m]_{n+1}=[m_{0,n+1}, m_{1,n+1}, \ldots , m_{n,n+1}],  
\label{mn+1}
\end{equation}
for which
\begin{equation}
m_{i,n+1}-m_{j,n+1}\in \Z_+ \quad (1\leq i< j \leq n). 
\label{cond}
\end{equation}
In a standard weight space basis, the highest weight $\Lambda$ of $W([m]_{n+1})$ is given
by 
\begin{equation}
\Lambda = m_{0,n+1}\, \epsilon + \sum_{i=1}^n m_{i,n+1}\, \delta_i .
\end{equation}

A Gel'fand-Zetlin basis for the $\gl(1|n)$ representation $W([m]_{n+1})$ has
been given and discussed in~\cite[Proposition~2]{KSV}, where it was shown that the set of vectors 
\begin{equation}
|m)_e = \left|
\begin{array}{lcllll}
m_{0,n+1} & m_{1,n+1}& \cdots & m_{n-2,n+1} & m_{n-1,n+1} & m_{n,n+1}  \\
& m_{1n} & \cdots & m_{n-2,n} & m_{n-1,n} & m_{nn}  \\
& m_{1,n-1} & \cdots & m_{n-2,n-1} &  m_{n-1,n-1}  &  \\
&\vdots & \qdots & & & \\
&m_{11} & & & &
\end{array} \right)_e
\label{m}
\end{equation}
satisfying the conditions
\begin{itemize}
\item[(GZ1)] $m_{i,n+1}$ are fixed and $m_{i,n+1}-m_{j,n+1}\in\Z_+$ \quad ($1\leq i< j\leq n$),
\item[(GZ2)] $m_{in}-m_{i,n+1}=\theta_{i}\in\{0,1\}$\quad ($1\leq i\leq n$),
\item[(GZ3)] if for $k\in\{1,\ldots,n\}$ one has $m_{0,n+1}+m_{k,n+1}=k-1$ then $\theta_k=0$, 
\item[(GZ4)] $m_{i,j+1}-m_{ij}\in\Z_+$ and $m_{ij}-m_{i+1,j+1}\in\Z_+$\quad ($1\leq i\leq j\leq n-1$),
\end{itemize}
constitute a basis in $W([m]_{n+1})$.
We have added here a subscript $e$ to the vectors $|m)_e$ in order to distinguish them
from another basis $|m)_E$ for $W([m]_{n+1})$ which will be introduced later.
For the explicit action of a set of $\gl(1|n)$ generators 
on the basis vectors~(\ref{m}), see~(\ref{e_00})-(\ref{fn}). 
Following~(\ref{mn+1}), it will be convenient to denote the elements of the other rows
in $|m)_e$, or more generally $k$-tuples, by
\begin{equation}
[m]_k=[m_{1k}, m_{2k}, \ldots, m_{kk}],\qquad (k=1,\ldots,n).
\end{equation}

With respect to the inner product $\langle |m')_e\,,\; |m)_e \rangle = \delta_{m,m'}$
and the condition $e_{ij}^\dagger=e_{ji}$, the representations $W([m]_{n+1})$ are unitary 
if and only if one of the following conditions is satisfied~\cite[Proposition~3]{KSV}:
\begin{itemize}
\item[(U1)] The highest weight is real and 
\begin{equation}
  m_{0,n+1}+m_{n,n+1}-n+1>0.
  \label{Utypical}
\end{equation}
In this case, the representation is typical.
\item[(U2)] The highest weight is real and there exists a $k\in \{ 1,2,\ldots, n\}$ such that
\begin{equation}
  m_{0,n+1}+m_{k,n+1}=k-1,\quad m_{k,n+1}=m_{k+1,n+1}=\cdots =m_{n,n+1}.
  \label{Uatypical}
\end{equation}
In this case, the representation is atypical of type~$k$.
\end{itemize}

Note that the highest weight vector of $W([m]_{n+1})$, denoted by $|\Lambda)_e$, is
given by:
\begin{equation}
|\Lambda)_e = \left|
\begin{array}{lcllll}
m_{0,n+1} & m_{1,n+1}& \cdots & m_{n-2,n+1} & m_{n-1,n+1} & m_{n,n+1}  \\
& m_{1,n+1} & \cdots & m_{n-2,n+1} & m_{n-1,n+1} & m_{n,n+1}  \\
& m_{1,n+1} & \cdots & m_{n-2,n+1} &  m_{n-1,n+1}  &  \\
&\vdots & \qdots & & & \\
&m_{1,n+1} & & & &
\end{array} \right)_e \ .
\label{hwv-e}
\end{equation}

Finally, note that the GZ-basis vectors $|m)_e$ are stationary states of
the quantum system. Indeed, by~(\ref{Haa}),~(\ref{eij}) and~(\ref{solution}) one has
\begin{equation}
\hat{H}=\hbar \left( (\sum_{j=1}^{n}\beta_j)\, e_{00} + \sum_{j=1}^n \beta_j \, e_{jj}\right),
\label{Hee}
\end{equation} 
so one finds, using~(\ref{e_00})-(\ref{e_kk}):
\begin{equation}
\hat{H} |m)_e = \hbar \tilde E_m |m)_e,
\end{equation}
with
\begin{equation}
\tilde E_m = (\sum_{j=1}^n \beta_j)(\sum_{l=0}^{n} m_{l,n+1}-\sum_{l=1}^n m_{ln}) + 
\sum_{j=1}^n \beta_j ( \sum_{l=1}^{j} m_{lj}-\sum_{l=1}^{j-1} m_{l,j-1}).
\end{equation}

\setcounter{equation}{0}
\section{Another set of $\gl(1|n)$ generators}
\label{sec:another}

The purpose is to describe the eigenvalues of $\q_r$ and $\p_r$, and to give their eigenvectors 
in terms of the GZ-basis vectors $|m)_e$ of $W([m]_{n+1})$, i.e.\ in terms of the stationary states.
The structure of $\q_r$ and $\p_r$ in terms of the generators $e_{0j}$ and $e_{j0}$ is 
similar, see~(\ref{q-e}) and~(\ref{p-e}), so it is sufficient to concentrate on
$\q_r$ only, with
\begin{equation}
\q_r  = \sqrt{\frac{\hbar}{\mu n}} \sum_{j=1}^n \left( \gamma_j\, e^{-2\pi i jr/n} e_{0j} +
\gamma_j\, e^{2\pi i jr/n} e_{j0} \right).
\label{q-e2}
\end{equation}
Since the description of $\q_r$ in terms of $e_{0j}$ and $e_{j0}$ is complicated (at least
for the action on GZ-basis vectors), we shall switch to another set of $\gl(1|n)$ generators.
For this purpose, recall the following proposition~\cite{KPSV}.
\begin{prop}
The Lie superalgebra generated by $2n$ odd elements
$\tilde e_{j0}$ and $\tilde e_{0j}$, with $1\leq j \leq n$, subject to the relations:
\begin{align}
& \{ \tilde e_{j0},\tilde e_{k0}\} = \{ \tilde e_{0j},\tilde e_{0k}\} = 0, \label{dr1}\\
&[\{ \tilde e_{j0},\tilde e_{0k}\}, \tilde e_{l0}] = \delta_{kl} \tilde e_{j0} - \delta_{jk} \tilde e_{l0}, \label{dr2}\\
&[\{ \tilde e_{j0},\tilde e_{0k}\}, \tilde e_{0l}] = \delta_{jk} \tilde e_{0l} - \delta_{lj} \tilde e_{0k}, \label{dr3}
\end{align}
is isomorphic to $\ssl(1|n)$.
\end{prop}
So clearly, our standard elements $e_{j0}$ and $e_{0j}$ generate $\ssl(1|n)$. The only difference between
$\ssl(1|n)$ and $\gl(1|n)$ comes from the Cartan subalgebra: for $\gl(1|n)$ this is spanned by
all elements $e_{jj}$ ($0\leq j\leq n$), and for $\ssl(1|n)$ by $e_{00}+e_{jj}$ ($1\leq j\leq n$).

The following proposition is easy but essential in our analysis:
\begin{prop}
Let  $U = (U_{jl})_{1\leq j\leq n,1\leq l\leq n}$ be a unitary $n\times n$ matrix,
and let
\begin{equation}
\label{def-E}
E_{j0}  = \sum_{l=1}^n U_{jl}\, e_{l0}\ \text{ and }\
E_{0j}  = \sum_{l=1}^n U_{jl}^*\, e_{0l}\qquad (1\leq j \leq n).
\end{equation}
Then the elements $E_{j0}$ and $E_{0j}$ satisfy the same defining relations 
(\ref{dr1})-(\ref{dr3}) as the elements $e_{j0}$ and $e_{0j}$. In other words,
also the $E_{j0}$ and $E_{0j}$ generate $\ssl(1|n)$.
\end{prop}
\noindent {\bf Proof.} 
It is a simple exercise to verify that
the elements $E_{j0}$ and $E_{0j}$ satisfy the relations (\ref{dr1})-(\ref{dr3}).
As an example, consider
\begin{equation*}
\begin{split}
[\{ E_{j0},E_{0k}\}, E_{l0}] & = 
\sum_{i_1,i_2,i_3} U_{ji_1}U^*_{ki_2}U_{li_3}[\{ e_{i_10},e_{0i_2}\}, e_{i_30}]\\
& = \sum_{i_1,i_2,i_3} U_{ji_1}U^*_{ki_2}U_{li_3}(\delta_{i_2i_3}e_{i_10} - 
\delta_{i_1i_2} e_{i_30}) \\
& = \sum_{i_2}U^*_{ki_2}U_{li_2}\sum_{i_1}U_{ji_1}e_{i_10} - 
\sum_{i_2}U_{ji_2}U^*_{ki_2}\sum_{i_3} U_{li_3}e_{i_30} \\
& = \sum_{i_2}U_{li_2}U^\dagger_{i_2k}\sum_{i_1}U_{ji_1}e_{i_10} - 
\sum_{i_2}U_{ji_2}U^\dagger_{i_2k}\sum_{i_3} U_{li_3}e_{i_30} \\
& = \delta_{lk} E_{j0} - \delta_{jk} E_{l0}.
\end{split}
\end{equation*}
\ \mybox

Using this proposition, it will be useful to identify the two parts of (\ref{q-e2}),
$\sum_{j=1}^n \gamma_j\, e^{-2\pi i jr/n} e_{0j}$ and $\sum_{j=1}^n \gamma_j\, e^{2\pi i jr/n} e_{j0}$ 
as single generators $E_{0k}$ and $E_{k0}$ for some $k$. Hence, let us define:
\begin{align}
E_{n0} & = \frac{1}{\sqrt{\gamma_1^2+\cdots+\gamma_n^2}}
\sum_{j=1}^n \gamma_j\, e^{2\pi i j r/n}e_{j0} 
= \frac{1}{\sqrt{\gamma}} \sum_{j=1}^n \gamma_j\, e^{2\pi i j r/n}e_{j0},
\label{En0} \\
E_{0n} & = \frac{1}{\sqrt{\gamma_1^2+\cdots+\gamma_n^2}}
\sum_{j=1}^n \gamma_j\, e^{-2\pi i j r/n}e_{0j} 
= \frac{1}{\sqrt{\gamma}} \sum_{j=1}^n \gamma_j\, e^{-2\pi i j r/n}e_{0j}.
\label{E0n}
\end{align}
Note that we have divided by $\sqrt{\gamma}$, so that the coefficients are entries
of a unitary matrix $U$, as required in~(\ref{def-E}). 
Next, we should supplement (\ref{En0}) and (\ref{E0n}) by other linear combinations of
the $e_{j0}$ and $e_{0j}$, such that the transition matrix is unitary. 
In principle, any matrix $U$ with last row $U_{nj}=\gamma_j\, e^{2\pi i j r/n}/\sqrt{\gamma}$ 
could be proposed.
However, in order to make computations for eigenvectors easier, we
will propose a matrix $U$ that is as simple as possible, i.e.\ with as many zero entries as
possible. Note that one cannot make a triangular choice for $U$, since the only triangular matrix
that is also unitary is diagonal. So we will make a choice that is as close as possible to
a triangular matrix, namely a Hessenberg matrix, so that all entries $U_{jl}$ with $l>j+1$ are zero.
This leads to the following expressions, for $j=1,2,\ldots,n-1$:
\begin{align}
\label{Ej0}
E_{j0} & = \frac{1}{ \sqrt{\frac{1}{\gamma_1^2+\cdots+ \gamma_{j}^2} 
+ \frac{1}{\gamma_{j+1}^2}} } 
\left( \sum_{l=1}^{j} 
\frac{e^{2\pi i r l/n}}
     {\gamma_1^2+\cdots + \gamma_{j}^2} \gamma_l e_{l0} 
     - \frac{1}{\gamma_{j+1}} e^{2\pi i r (j+1)/n} e_{j+1,0}\right),\\
\label{E0j}
E_{0j} & = \frac{1}{ \sqrt{\frac{1}{\gamma_1^2+\cdots+ \gamma_{j}^2} 
+ \frac{1}{\gamma_{j+1}^2}} } 
\left( \sum_{l=1}^{j} 
\frac{e^{-2\pi i r l/n}}
     {\gamma_1^2+\cdots + \gamma_{j}^2} \gamma_l e_{0l} 
     - \frac{1}{\gamma_{j+1}} e^{-2\pi i r (j+1)/n} e_{0,j+1}\right).     
\end{align}
It is a simple exercise to verify that the transition matrix $U$ defined by means of (\ref{Ej0})
and (\ref{En0}) is indeed a unitary matrix.
So the operators (\ref{En0})-(\ref{E0j}) form a set of generators for $\ssl(1|n)$,
such that the position operator $\q_r$ becomes:
\begin{equation}
\q_r=  \sqrt{\frac{\hbar\gamma}{\mu n}} (E_{0n}+E_{n0}).
\label{q-E}
\end{equation}
Note that for every different position operator $\q_r$ (i.e.\ for every different $r$), 
one has a {\em different set of generators}, so we should denote them by $E_{j0}^{(r)}$ and
$E_{0j}^{(r)}$. This overloads the notation, however. So we shall assume that $r$ is
fixed, and drop the superscript $(r)$ from the generators.

The elements $E_{0j}$ and $E_{j0}$ generate $\ssl(1|n)$. The new odd basis elements of $\ssl(1|n)$ are
directly given by (\ref{En0})-(\ref{E0j}). The new even basis elements of $\ssl(1|n)$
are of the form $E_{jk}=\{E_{j0},E_{0k}\}$ with $j\ne k$, and $\{E_{j0},E_{0j}\}$.
Since, without writing the matrix elements of $U$ explicitly as in~(\ref{En0})-(\ref{E0j}), 
for $j=1,2,\ldots,n$:
\begin{equation}
\{ E_{j0},E_{0j}\} = e_{00} + \sum_{l=1}^n \sum_{k=1}^n U_{jl}U^*_{jk}\; e_{lk}
\label{EjEj}
\end{equation}
one can extend the new $\ssl(1|n)$ basis to a $\gl(1|n)$ basis by putting $E_{00}=e_{00}$,
and $E_{jj}$ equal to the remaining part in (\ref{EjEj}), i.e.\ $E_{jj}=\{E_{j0},E_{0j}\}-e_{00}$.

So we have a {\em new basis} $E_{ij}$ for $\gl(1|n)$, satisfying the same relations~(\ref{eij})
as the old basis $e_{ij}$, and the same star conditions $E_{ij}^\dagger=E_{ji}$. 
In terms of this new basis, the position operator $\q_r$ has a simple expression, see~(\ref{q-E}).
Due to this simple expression, the eigenvalues and eigenvectors of $\q_r$ can be computed.
With respect to this new basis, the representation $W([m]_{n+1})$ has a new highest weight
vector, to be denoted by $|\Lambda)_E$. 
One essential task will be the expansion of $|\Lambda)_E$ in terms of the old GZ-basis vectors
$|m)_e$.
Also with respect to this new basis $E_{ij}$, one can define a new GZ-basis for $W([m]_{n+1})$,
the vectors of this basis being denoted by $|m)_E$. The action of $E_{ij}$ on vectors $|m)_E$ is 
identical to the action of $e_{ij}$ on vectors $|m)_e$. 

In the following section we shall consider the branching $\gl(1|n)\rightarrow \gl(1|1)\oplus \gl(n-1)$
for $W([m]_{n+1})$, with respect to this new basis. This will yield the eigenvalues of 
$\q_r$ in $W([m]_{n+1})$. The (orthonormal) eigenvectors of $\q_r$ are easy to 
describe in the $|m)_E$ basis of $W([m]_{n+1})$.
In Section~\ref{sec:vectors}, we make the connection between the old basis vectors $|m)_e$ and
the new ones $|m)_E$, leading to a description of the $\q_r$ eigenvectors in
the original basis.

\setcounter{equation}{0}
\section{On the decomposition $\gl(1|n)\rightarrow \gl(1|1)\oplus \gl(n-1)$ for unitary representations}
\label{sec:decomposition}

Consider the Lie superalgebra $\gl(1|n)$ with (new) basis elements 
$E_{ij}$ ($i,j=0,1,\ldots,n$) satisfying the standard relations~(\ref{eij}).
We consider the finite-dimensional unitary irreducible representations $W([m]_{n+1})$
with GZ-basis vectors $|m)_E$. 
The action of $E_{ij}$ on $|m)_E$ is identical to that of $e_{ij}$ on $|m)_e$
(see~(\ref{e_00})-(\ref{fn})). In particular,
the diagonal action reads:
\begin{equation}
\begin{split}
E_{00}|m)_E& =\bigl(m_{0,n+1}-\sum_{j=1}^n \theta_{j}\bigr)|m)_E; \\
E_{jj}|m)_E& =\bigl(\sum_{l=1}^j m_{lj}-\sum_{l=1}^{j-1} m_{l,j-1}\bigr)|m)_E, 
\quad (1\leq j\leq n).
\label{E_jj}
\end{split}
\end{equation}

In order to describe the decomposition $\gl(1|n)\rightarrow \gl(1|1)\oplus \gl(n-1)$
for such unitary representations, let us first list (and fix the notation for) the
unitary representations of $\gl(1|1)$~\cite{KSV}.
Let $\{e_{00},e_{10},e_{01}, e_{11}\}$
be a basis for $\gl(1|1)$, and denote the highest weight labels $[m_{0,2},m_{1,2}]$ by $[a,b]$
and the representation itself by $W([a,b])$. Then, following (U1)-(U2), there are two possibilities:
\begin{itemize}
\item[(1)] A typical unitary $\gl(1|1)$ representation $W([a,b])$, with $a,b\in\R$ and $a+b>0$. The GZ-basis of the 
representation consists of two vectors only, which we shall denote by $v$ and $w$, and the
action is given by:
\begin{equation}
\begin{array}{ll}
e_{00}\,v = a\,v, &e_{00}\,w=(a-1)\,w,\\
e_{11}\,v = b\,v,  &e_{11}\,w=(b+1)\,w,\\
e_{01}\,v = 0,  &e_{01}\,w=\sqrt{a+b}\;v,\\
e_{10}\,v = \sqrt{a+b}\;w, & e_{10}\,w=0.
\end{array}
\label{action-v-w}
\end{equation}
The weights of the representation are $(a,b)$ and $(a-1,b+1)$.
\item[(2)] An atypical unitary $\gl(1|1)$ representation $W([a,b])$, with $a,b\in\R$ and $a+b=0$. The GZ-basis 
consists of one vector only, denoted by $v$, and the only non-zero actions are
\begin{equation}
e_{00}\,v = a\,v, \qquad e_{11}\,v=-a\,v.
\end{equation}
The weight of the representation is $(a,-a)$.
\end{itemize}

The new GZ-basis (and the new basis $E_{ij}$ for $\gl(1|n)$) 
can now be used to find the decomposition $\gl(1|n)\rightarrow \gl(1|1)\oplus \gl(n-1)$
for unitary $\gl(1|n)$ representations. For this purpose, it is convenient to take
\begin{equation}
\{E_{00},E_{n0},E_{0n}, E_{nn}\}
\label{gl11-basis}
\end{equation}
as the basis elements of $\gl(1|1)$, and 
$\{E_{ij} | 1\leq i,j \leq n-1\}$ as the basis elements of $\gl(n-1)$. Indeed,
with this choice the actions of the $\gl(1|1)$ generators on $|m)_E$ only change the labels
in the second row of the GZ-pattern (see~(\ref{en})-(\ref{fn})); 
and the actions of the $\gl(n-1)$ generators only change the last $(n-1)$ rows~(see (\ref{ek})-(\ref{fk}) 
for $2\leq k\leq n-1$).
Otherwise said, the last $(n-1)$ rows of $|m)_E$ coincide with the usual $\gl(n-1)$ GZ-basis labels.
Note that the action of the diagonal elements of $\gl(1|1)$ is given by
\begin{equation}
E_{00}|m)_E=a\;|m)_E, \qquad
E_{nn}|m)_E= b\;|m)_E, 
\end{equation}
with
\begin{align}
& a= \sum_{j=0}^n m_{j,n+1}-\sum_{j=1}^n m_{jn} = m_{0,n+1}-\sum_{j=1}^n \theta_{j}, \\
& b= \sum_{j=1}^n m_{jn}-\sum_{j=1}^{n-1} m_{j,n-1}.
\end{align}
For a given unitary representation $W([m]_{n+1})$ of $\gl(1|n)$, the decomposition 
to $\gl(1|1)\oplus \gl(n-1)$ is thus completely determined by listing all possible
rows $[m]_n$ and $[m]_{n-1}$, i.e.
\begin{equation}
\begin{array}{lcll}
m_{1n}=m_{1,n+1}+\theta_1,  & \cdots &m_{n-1,n}=m_{n-1,n+1}+\theta_{n-1}, & m_{nn}=m_{n,n+1}+\theta_n \\
m_{1,n-1},  & \cdots & m_{n-1,n-1} & 
\end{array}
\label{2rows}
\end{equation}
subject to conditions (GZ3) and (GZ4), with $\theta_i\in\{0,1\}$. 

Let us investigate the $\gl(1|1)$ weight $(a,b)$ of a vector $|m)_E$ more carefully.
For a {\em typical} unitary representation $W([m]_{n+1})$, one finds
\begin{equation*}
\begin{split}
a+b &= \sum_{j=0}^n m_{j,n+1}-\sum_{j=1}^{n-1} m_{j,n-1}
 = m_{0,n+1}+m_{n,n+1}+\sum_{j=1}^{n-1}(m_{j,n+1}-m_{j,n-1}) \\
 &= m_{0,n+1}+m_{n,n+1}+\sum_{j=1}^{n-1}(m_{jn}-m_{j,n-1})- \sum_{j=1}^{n-1} \theta_j.
\end{split}
\end{equation*}
But by (GZ4) $m_{jn}-m_{j,n-1}\geq 0$, and by (\ref{Utypical}) $m_{0,n+1}+m_{n,n+1}>n-1$,
hence
\begin{equation*}
a+b > n-1 - \sum_{j=1}^{n-1} \theta_j \geq 0,
\end{equation*}
so $a+b>0$ and $(a,b)$ can be the weight of a typical 2-dimensional $\gl(1|1)$ representation only.

For an {\em atypical} unitary representation $W([m]_{n+1})$, satisfying~(\ref{Uatypical}), (GZ2)-(GZ4)
imply that $\theta_k=\theta_{k+1}=\cdots=\theta_n=0$. Then
\begin{equation*}
\begin{split}
a+b &= m_{0,n+1}+m_{n,n+1}+\sum_{j=1}^{n-1}(m_{jn}-m_{j,n-1})- \sum_{j=1}^{n-1} \theta_j \\
&= m_{0,n+1}+m_{k,n+1}+\sum_{j=1}^{n-1}(m_{jn}-m_{j,n-1})- \sum_{j=1}^{k-1} \theta_j \\
&= (k-1)- \sum_{j=1}^{k-1} \theta_j +\sum_{j=1}^{n-1}(m_{jn}-m_{j,n-1}). \\
\end{split}
\end{equation*}
So $a+b\geq 0$, and $a+b$ can be equal to zero if and only if
\begin{equation}
\begin{split}
& \theta_1=\theta_2=\cdots=\theta_{k-1}=1, \hbox{ and }\\
& m_{j,n-1}=m_{jn} \hbox{ for all } j=1,2,\ldots,n-1.
\end{split}
\end{equation}
Consequently, in the $\gl(1|n)\rightarrow \gl(1|1)\oplus \gl(n-1)$ decomposition
\begin{equation}
W([m]_{n+1}) \rightarrow \bigoplus\ W([a,b]) \times V([m]_{n-1})
\end{equation}
the representations $W([a,b])$ are always typical if $W([m]_{n+1})$ is typical.
If $W([m]_{n+1})$ is atypical of type~$k$, again all $W([a,b])$ are typical, except
for one single component where $a+b=0$ and where the labels of the $\gl(n-1)$ representation $V([m]_{n-1})$ 
are given by:
\begin{equation}
\begin{split}
& m_{j,n-1}=m_{jn}=m_{j,n+1}+1 \hbox{ for } j=1,\ldots,k-1;\\
& m_{j,n-1}=m_{jn}=m_{j,n+1} \hbox{ for } j=k,\ldots,n-1.
\end{split}
\label{a+b=0}
\end{equation}

Following~(\ref{2rows}), for a fixed $(n+1)$-tuple $[m]_{n+1}$ satisfying (\ref{cond}), the set of $(n-1)$-tuples
appearing in the decomposition to $\gl(n-1)$ is given by
\begin{equation}
\begin{split}
{\cal M}([m]_{n+1}) &= \{ [m]_{n-1}\;|\; m_{i,n+1}+1-m_{i,n-1},\; m_{i,n-1}-m_{i+1,n+1}\in\Z_+,\ 
(i=1,\ldots,n-1); \\
& \qquad m_{i,n-1}-m_{i+1,n-1}\in\Z_+,\ 
(i=1,\ldots,n-2) \}.
\end{split}
\end{equation}
To see how often such a $\gl(n-1)$ representation labeled by $[m]_{n-1}$ appears in
the decomposition of $W([m]_{n+1})$, one should count the number of allowed $\theta_i$'s
in~(\ref{2rows}). Since each $\theta_i\in\{0,1\}$, this number will be a power of~2. Whether
both values for $\theta_i$ are allowed depends not only on the (a)typicality of $W([m]_{n+1})$,
but also on the tuples $[m]_{n+1}$ and $[m]_{n-1}$ themselves (whether some consecutive numbers 
are equal, whether some $m_{i,n-1}$ is equal to $m_{i,n+1}+1$, etc.). 
Both values for $\theta_i$ are allowed if $m_{i-1,n-1}-m_{i,n+1}-1\in\Z_+$ and 
$m_{i,n+1}-m_{i,n-1}\in\Z_+$ (for $i=1$ the first condition disappears since $m_{0,n-1}$
is not a GZ-label, and for $i=n$ the second condition disappears since $m_{n,n-1}$ is
not a GZ-label). So let us consider
\begin{equation}
T(m_{i-1,n-1}-m_{i,n+1}-1\in\Z_+ \hbox{ and }m_{i,n+1}-m_{i,n-1}\in\Z_+)
\end{equation}
where $T(\hbox{A})=1$ if A is true and $T(\hbox{A})=0$ if A is false, and
\begin{equation}
N([m]_{n+1},[m]_{n-1}) = \sum_{i=1}^n 
T(m_{i-1,n-1}-m_{i,n+1}-1\in\Z_+ \hbox{ and }m_{i,n+1}-m_{i,n-1}\in\Z_+).
\label{N}
\end{equation}
Then, for a typical $\gl(1|n)$ representation $W([m]_{n+1})$, the number of $\gl(n-1)$
representations $V([m]_{n-1})$ appearing in the decomposition (with $[m]_{n-1} \in {\cal M}([m]_{n+1})$)
is given by 
\begin{equation}
2^{N([m]_{n+1},[m]_{n-1})}.
\label{2^N}
\end{equation}
For a $\gl(1|n)$ representation that is atypical of type~$k$, the result is essentially the same
but now all $\theta_k=\cdots=\theta_n=0$. So in this case the result is still given by~(\ref{2^N}),
except that the upper bound of the sum in~(\ref{N}) is $k-1$ instead of $n$.
It will be convenient to have a notation for the set of allowed $n$-tuples, for a given $(n+1)$-tuple
$[m]_{n+1}$ and a given $(n-1)$-tuple $[m]_{n-1}$:
\begin{equation}
{\cal A}([m]_{n+1},[m]_{n-1})= \{ [m]_n \,|\, [m]_{n+1}, [m]_{n}\hbox{ and }[m]_{n-1}\hbox{ satisfy (GZ2)-(GZ4)}\}.
\end{equation}
So the number of elements of ${\cal A}([m]_{n+1},[m]_{n-1})$ is given by~(\ref{2^N}).

Knowing the multiplicity of $V([m]_{n-1})$, one can now determine the $\gl(1|1)$ weights $(a,b)$ 
for each appearance of $V([m]_{n-1})$ in the decomposition of $W([m]_{n+1})$, and collect these
according to irreducible representations of $\gl(1|1)$ (which are one- or two-dimensional). 
This gives rise to the following:
\begin{equation}
W([m]_{n+1}) \rightarrow \bigoplus_{[m]_{n-1}\in {\cal M}([m]_{n+1}) }
\left( \bigoplus_{i=0}^{N-1} \binom{N-1}{i} W([a-i,b+i]) \right)\times V([m]_{n-1})
\label{branching}
\end{equation}
where
\begin{equation*}
\begin{split}
N &\equiv N([m]_{n+1},[m]_{n-1}), \\
a &= \sum_{j=0}^{n} m_{j,n+1} - \min_{[m]_n\in {\cal A}([m]_{n+1},[m]_{n-1})} \left( \sum_{j=1}^n m_{jn}\right),\\
b &= -a + \sum_{j=0}^n m_{j,n+1}-\sum_{j=1}^{n-1} m_{j,n-1}.
\end{split}
\end{equation*}
Note that for typical representations each $N>0$. For representations atypical of type~$k$,
there is one single $(n-1)$-tuple $[m]_{n-1}$ for which $N=N([m]_{n+1},[m]_{n-1})=0$, namely the
case~(\ref{a+b=0}). For this $(n-1)$-tuple, the term in the right hand side of~(\ref{branching})
should be replaced by
\begin{equation}
W([a,-a]) \times V([m]_{n-1}).
\end{equation}

It will be important to notice that the range of values for $a+b$
in $W([m]_{n+1})$ goes in steps of~1 and follows from~(\ref{branching}); it is given by
\begin{equation}
m_{0,n+1}+m_{1,n+1}, m_{0,n+1}+m_{1,n+1}-1,  \ldots, m_{0,n+1}+m_{n,n+1}-n+1\ (>0)
\end{equation}
for typical representations, and by
\begin{equation}
m_{0,n+1}+m_{1,n+1}, m_{0,n+1}+m_{1,n+1}-1,  \ldots, m_{0,n+1}+m_{k,n+1}-k+1\ (=0)
\end{equation}
for representations atypical of type~$k$.

We are now in a position to solve the eigenvalue problem for $\q_r$. Remember
that $\q_r= \sqrt{\frac{\hbar\gamma}{\mu n}}(E_{0n}+E_{n0})$, see~(\ref{q-E}).
Hence in a 2-dimensional typical $\gl(1|1)$ representation $W([a,b])$ ($a+b>0$), it follows
from~(\ref{action-v-w}) and~(\ref{gl11-basis}) that the eigenvalues of $E_{0n}+E_{n0}$ are
$\pm\sqrt{a+b}$, whereas in a 1-dimensional atypical $\gl(1|1)$ representation $W([a,b])$
($a+b=0$), the eigenvalue is~$0$. 

So we find the following result:
\begin{theo}
Let $W([m]_{n+1})$ be a unitary representation of $\gl(1|n)$.
\begin{itemize}
\item[(a)] If $W([m]_{n+1})$ is typical, the eigenvalues of $\q_r$ are given by 
$\pm\sqrt{\frac{\hbar\gamma K}{\mu n}}$ where the range of $K$, in steps of~1, is determined by
\begin{equation}
K=m_{0,n+1}+m_{1,n+1}, m_{0,n+1}+m_{1,n+1}-1,  \ldots, m_{0,n+1}+m_{n,n+1}-n+1.
\end{equation}
The multiplicity of each eigenvalue $\pm\sqrt{\frac{\hbar\gamma K}{\mu n}}$ 
is determined by~(\ref{branching}) and is of the form
\begin{equation}
\sum 2^N \dim(V([m]_{n-1}))
\label{multiplicity}
\end{equation}
where the sum is over all $(n-1)$-tuples $[m]_{n-1}$ from ${\cal M}([m]_{n+1})$ for which
$\sum_{j=0}^n m_{j,n+1}-\sum_{j=1}^{n-1} m_{j,n-1} =K$.
The dimensions of $\gl(n-1)$ representations $V([m]_{n-1})$ are well known~\cite[p.~33]{Wybourne}.
\item[(b)] If $W([m]_{n+1})$ is atypical of type~$k$, the eigenvalues of $\q_r$ are given by 
$\pm\sqrt{\frac{\hbar\gamma K}{\mu n}}$ where $K=0,1,2,\ldots,m_{0,n+1}+m_{1,n+1}$.
The multiplicity of each nonzero eigenvalue is again determined by~(\ref{branching}) and 
given by a formula similar to~(\ref{multiplicity}).
The multiplicity of the zero eigenvalue is $\dim V([m]_{n-1})$,
with $[m]_{n-1}$ given by~(\ref{a+b=0}).
\end{itemize}
\end{theo}

\setcounter{equation}{0}
\section{Relation between the two GZ-basis vectors}
\label{sec:vectors}

Consider the unitary $\gl(1|n)$ representation $W([m]_{n+1})$. 
On the one hand, $W([m]_{n+1})$ has a GZ-basis of vectors $|m)_e$, with the
standard action of $e_{ij}$ on these vectors determined by~(\ref{e_00})-(\ref{ej0}).
The highest weight vector $|\Lambda)_e$ with respect to this $\gl(1|n)$ basis is given by~(\ref{hwv-e}).
Note that the highest weight vector is uniquely characterized by:
\begin{align}
& e_{j,j+1}\, |\Lambda)_e = 0 \qquad (1\leq j \leq n-1), \label{even-hw}\\
& e_{0j}\, |\Lambda)_e = 0 \qquad (1\leq j \leq n). \label{odd-hw}
\end{align}
The last condition is guaranteed by the fact that for $|\Lambda)_e$ all $\theta_i=0$ in~(GZ2).
The first condition follows from the action~(\ref{ek}).

On the other hand, we have considered a new basis $E_{ij}$ for $\gl(1|n)$, determined by (\ref{En0})-(\ref{E0j}).
With respect to this new basis, $W([m]_{n+1})$ has a new GZ-basis with vectors $|m)_E$, 
and a new highest weight vector $|\Lambda)_E$. 
We want to find an expression for $|\Lambda)_E$ as a linear combination of vectors $|m)_e$:
\begin{equation}
|\Lambda)_E = \sum c_m\; |m)_e.
\label{E-as-e}
\end{equation}
So, we should require:
\begin{align}
& E_{j,j+1}\, |\Lambda)_E = 0 \qquad (1\leq j \leq n-1), \label{E-even-hw}\\
& E_{0j}\, |\Lambda)_E = 0 \qquad (1\leq j \leq n). \label{E-odd-hw}
\end{align}
Since each $E_{0j}$ is a linear combination of elements $e_{0l}$, it follows that the
linear combination in~(\ref{E-as-e}) consists of $m$-patterns with all $\theta_i=0$ in~(GZ2).
So we should examine the elements $E_{j,j+1}$ more closely, and in particular their
action on vectors $|m)_e$. 

We can compute $E_{j,j+1}$ by means of (\ref{Ej0})-(\ref{E0j}) and $E_{j,j+1}=\{E_{j0},E_{0,j+1}\}$.
For $1\leq j\leq n-2$, this gives
\begin{equation}
\begin{split}
& E_{j,j+1} =  \gamma_{j+1}\gamma_{j+2}
\sqrt{ \frac{\gamma_1^2+\cdots+ \gamma_{j}^2} 
{\gamma_1^2+\cdots + \gamma_{j+2}^2}  } \
\Bigl( \sum_{l_1=1}^{j}\sum_{l_2=1}^{j+1} 
\frac{ e^{2\pi i r(l_1-l_2)/n} \gamma_{l_1}\gamma_{l_2}}
{(\gamma_1^2+\cdots+\gamma_{j}^2)(\gamma_1^2+\cdots + \gamma_{j+1}^2)}
e_{l_1l_2} \\
&\quad -\sum_{l_1=1}^{j} \frac{e^{2\pi i r(l_1-j-2)/n} \gamma_{l_1}  }
{(\gamma_1^2+\cdots+\gamma_{j}^2)\gamma_{j+2}}e_{l_1,j+2}
-\sum_{l_2=1}^{j+1} \frac{e^{2\pi i r(j+1-l_2)/n} \gamma_{l_2}  }
{(\gamma_1^2+\cdots+\gamma_{j+1}^2)\gamma_{j+1}}e_{j+1,l_2} 
+\frac{e^{-2\pi i r/n}  }{\gamma_{j+1}\gamma_{j+2}} e_{j+1,j+2}\Bigr),
\end{split}
\label{Ej-e}
\end{equation}
and for $j=n-1$:
\begin{equation}
E_{n-1,n}   = 
\frac{ \gamma_n \sqrt{\gamma_1^2+\cdots+ \gamma_{n-1}^2}} 
{\gamma_1^2+\cdots + \gamma_{n}^2}   \
\Bigl( \sum_{l_1=1}^{n-1}\sum_{l_2=1}^{n} 
\frac{ e^{2\pi i r(l_1-l_2)/n} \gamma_{l_1}\gamma_{l_2}}
{(\gamma_1^2+\cdots+\gamma_{n-1}^2)}
e_{l_1l_2} 
-\sum_{l=1}^{n} \frac{e^{-2\pi i rl/n} \gamma_{l}  }
{\gamma_n}e_{nl}\Bigr).
\label{En-e}
\end{equation}

The following type of vectors from $W([m]_{n+1})$ will play an essential role:
\begin{equation}
|m(d))_e = \left|
\begin{array}{lcllllll}
m_{0,n+1} & m_{1,n+1}& m_{2,n+1} & \cdots &\cdots & m_{n-2,n+1} & m_{n-1,n+1} & m_{n,n+1}  \\
& m_{1,n+1} & m_{2,n+1} & \cdots &\cdots & m_{n-2,n+1} & m_{n-1,n+1} & m_{n,n+1}  \\
& m_{1,n+1} & m_{2,n+1} & \cdots &\cdots & m_{n-2,n+1} &  m_{n-1,n-1}  &  \\
&\vdots & \vdots & \vdots& \vdots &\qdots & \\
&m_{1,n+1} & m_{2,n+1} & m_{3, n+1}& m_{44}&& \\
&m_{1,n+1} & m_{2,n+1} & m_{33}&&& \\
&m_{1,n+1}  & m_{22}&&&& \\
&m_{11} & & & & &
\end{array}
\right)_e
\label{md}
\end{equation}
So in this expression, all labels in the GZ-pattern are fixed, except the $(n-1)$ bottom
labels $d=(m_{11},m_{22},\ldots,m_{n-1,n-1})$ which are allowed to vary according
to~(GZ4). 

Now we have the following result.
\begin{prop}
The highest weight vector of $W([m]_{n+1})$ according to the new $\gl(1|n)$ basis $E_{ij}$
is given by:
\begin{align}
|\Lambda)_E & =\frac{1}{\sqrt{\cal N}}\sum_{m_{n-1,n-1}=m_{n,n+1}}^{m_{n-1,n+1}} \; 
\sum_{m_{n-2,n-2}=m_{n-1,n-1}}^{m_{n-2,n+1}} \cdots 
\sum_{m_{22}=m_{33}}^{m_{2,n+1}} \;
\sum_{m_{11}=m_{22}}^{m_{1,n+1}} (-1)^{m_{11}+\cdots +m_{n-1,n-1}}\nn \\
& \times e^{-2\pi i r(m_{11}+ \cdots +m_{n-1,n-1})/n}
\left[ \binom{m_{1,n+1}-m_{22}}{m_{1,n+1}-m_{11}}   
\binom{m_{2,n+1}-m_{33}}{m_{2,n+1}-m_{22}} \cdots \right. \nn \\
& \times \left. \cdots \binom{m_{n-2,n+1}-m_{n-1,n-1}}{m_{n-2,n+1}-m_{n-2,n-2}}  
\binom{m_{n-1,n+1}-m_{n,n+1}}{m_{n-1,n+1}-m_{n-1,n-1}}   \right]^{1/2} \nn\\
& \times \gamma_1^{m_{1,n+1}-m_{11}} \gamma_2^{m_{11}-m_{22}} \gamma_3^{m_{22}-m_{33}} \ldots 
\gamma_{n-1}^{m_{n-2,n-2}-m_{n-1,n-1}} 
\gamma_{n}^{m_{n-1,n-1}-m_{n,n+1}} |m(d))_e,
\label{E-hw}
\end{align}
where ${\cal N}$ is a normalization factor given by:
\begin{equation}
{\cal N}=(\gamma_1^2+\gamma_2^2)^{m_{1,n+1}-m_{2,n+1}} 
(\gamma_1^2+\gamma_2^2+\gamma_3^2)^{m_{2,n+1}-m_{3,n+1}} \cdots
(\gamma_1^2+\cdots +\gamma_n^2)^{m_{n-1,n+1}-m_{n,n+1}} .
\label{calN}
\end{equation}
\end{prop}

\noindent {\bf Proof.} We shall only give a sketch of the proof, which requires careful computations.
Essentially, one considers for $1\leq j\leq n-1$ the action $E_{j,j+1}\,|\Lambda)_E$, 
using~(\ref{Ej-e})-(\ref{En-e}), (\ref{E-hw}) and the explicit action on the GZ-basis given 
by~(\ref{e_kk})-(\ref{fk}). In the resulting expression, one combines all contributions 
with the same GZ-pattern, and verifies that the coefficients become zero.
In this computation, it is essential to know the action of an element $e_{l_1l_2}$ on 
vectors of the form~(\ref{md}). 
From the general action~(\ref{e_kk})-(\ref{fk}), one deduces:
\begin{itemize}
\item If $l_1=l_2$, then $e_{l_1l_2} |m(d))_e$ gives just a constant times $|m(d))_e$.
\item If $l_1<l_2$, then $e_{l_1l_2} |m(d))_e$ gives only one term with a vector which is again of the 
form~(\ref{md}).
\item If $l_1>l_2$, then $e_{l_1l_2} |m(d))_e$ gives a linear combination of several vectors. Some of these
vectors are of the form~(\ref{md}). The other vectors are not of the form~(\ref{md}):
they have the same labels as $|m(d))_e$, but with
one of the labels in row~$l$ decreased by~1, for every $l=l_1-1,l_1-2,\ldots,l_2$. 
\end{itemize}
A careful examination shows that taking together all contributions 
to vectors that are not of the type~(\ref{md}) in the
expansion of $E_{j,j+1}\,|\Lambda)_E$ gives zero.
So it remains to compute the coefficients of vectors of the type~(\ref{md})
in the expansion of $E_{j,j+1}\,|\Lambda)_E$ ($j=1,\ldots,n-1$).
Explicitly, this gives rise to a coefficient of the form:
\begin{equation}
\frac{\prod_{i=1}^{j} (m_{i,n+1}-m_{ii})}{\prod_{i=1}^{j-1} (m_{i,n+1}-m_{i+1,i+1})} +
 \sum_{l=1}^{j-1}  (m_{ll}-m_{l+1,l+1})
 \frac{\prod_{i=l+1}^{j} (m_{i,n+1}-m_{ii})}{\prod_{i=l}^{j-1} (m_{i,n+1}-m_{i+1,i+1})}
 - m_{j,n+1}+m_{jj}. 
\label{coeff-md}
\end{equation}
Denote $m_{i,n+1}=x_i$ and $m_{ii}=y_i$. We shall prove that
\begin{equation}
\frac{\prod_{i=1}^{j} (x_{i}-y_{i})}{\prod_{i=1}^{j-1} (x_{i}-y_{i+1})} +
 \sum_{l=1}^{j-1} (y_l-y_{l+1}) 
 \frac{\prod_{i=l+1}^{j} (x_{i}-y_{i})}{\prod_{i=l}^{j-1} (x_{i}-y_{i+1})}
 = x_{j}-y_{j}
\label{id-xy} 
\end{equation}
for arbitrary variables $x_i$ and $y_i$, implying that (\ref{coeff-md}) is indeed always zero. 
The identity~(\ref{id-xy}) is true for $j=1$. 
Suppose it is true for a fixed $j$, and let us consider it 
for $j+1$:
\begin{equation}
\frac{\prod_{i=1}^{j+1} (x_{i}-y_{i})}{\prod_{i=1}^{j} (x_{i}-y_{i+1})} +
 \sum_{l=1}^{j} (y_l-y_{l+1}) 
 \frac{\prod_{i=l+1}^{j+1} (x_{i}-y_{i})}{\prod_{i=l}^{j} (x_{i}-y_{i+1})}
 = x_{j+1}-y_{j+1}. 
\label{id-xy+}
\end{equation}
The left hand side of (\ref{id-xy+}) yields, using~(\ref{id-xy}) and induction on~$j$:
\begin{equation*}
\begin{split}
& \frac{(x_{j+1}-y_{j+1})}{(x_{j}-y_{j+1})}\Big( 
\frac{\prod_{i=1}^{j} (x_{i}-y_{i})}{\prod_{i=1}^{j-1} (x_{i}-y_{i+1})} +
 \sum_{l=1}^{j-1} (y_l-y_{l+1}) 
 \frac{\prod_{i=l+1}^{j} (x_{i}-y_{i})}{\prod_{i=l}^{j-1} (x_{i}-y_{i+1})}\Big)
 + (y_{j}-y_{j+1})\frac{(x_{j+1}-y_{j+1})}{(x_{j}-y_{j+1})}\\
&\qquad = \frac{(x_{j+1}-y_{j+1})}{(x_{j}-y_{j+1})}\Big(x_{j}-y_{j}\Big) 
 +  (y_{j}-y_{j+1})\frac{(x_{j+1}-y_{j+1})}{(x_{j}-y_{j+1})}
 = x_{j+1}-y_{j+1}.
\end{split} 
\end{equation*}
So the identity holds in general. This shows that all coefficients in the expansion
of $E_{j,j+1}\,|\Lambda)_E$ are zero, in other words $E_{j,j+1}\,|\Lambda)_E=0$. 

To see that ${\cal N}$ gives the right normalization coefficient, one can simply
expand the right hand side of~(\ref{calN}). This gives, after appropriate relabeling
of the summation indices:
\begin{align}
&\sum_{k_{n-1}=m_{n,n+1}}^{m_{n-1,n+1}} \; 
\sum_{k_{n-2}=k_{n-1}}^{m_{n-2,n+1}} \cdots 
\sum_{k_{2}=k_{3}}^{m_{2,n+1}} \;
\sum_{k_{1}=k_{2}}^{m_{1,n+1}} 
\binom{m_{1,n+1}-k_{2}}{m_{1,n+1}-k_{1}}   
\binom{m_{2,n+1}-k_{3}}{m_{2,n+1}-k_{2}} \cdots \\
& \cdots \binom{m_{n-1,n+1}-m_{n,n+1}}{m_{n-1,n+1}-k_{n-1}}  
\; \gamma_1^{2(m_{1,n+1}-k_{1})} \gamma_2^{2(k_{1}-k_{2})} \gamma_3^{2(k_{2}-k_{3})} \ldots 
\gamma_{n}^{2(k_{n-1}-m_{n,n+1})}.
\end{align}
Clearly, this is just the norm of the vector given as a summand in the right hand side
of~(\ref{E-hw}).
\mybox

In principle, we now have a solution to our eigenvector problem, i.e.\ we can
give a set of orthonormal eigenvectors of $\q_r$ for $W([m]_{n+1})$ in terms of
the basis $|m)_e$.
First of all, (\ref{branching}) gives the decomposition of $W([m]_{n+1})$ with respect to
$\gl(1|1)\oplus\gl(n-1)$, so from this step one can express the weight vectors $v$ and $w$
of every $W([a,b])$ ($a+b>0$) in terms of vectors $|m)_E$. 
Then~(\ref{action-v-w}) and~(\ref{q-E}) imply that the eigenvectors of $\q_r$
are $(v\pm w)/\sqrt{2}$:
\begin{equation}
\q_r \frac{v\pm w}{\sqrt{2}} = \pm \sqrt{\frac{\hbar\gamma}{\mu n}}\sqrt{a+b}\ \frac{v\pm w}{\sqrt{2}}.
\label{qr-vw}
\end{equation}
But in principle every $|m)_E$, and thus also $v$ and $w$, can be expressed as powers of $E_{ij}$ ($i>j$)
acting on $|\Lambda)_E$ (in practice this can be hard, though). 
The rest is now routine: write every such $E_{ij}$ in terms of $e_{ij}$, and use~(\ref{E-hw}).
This leads to an expression of the eigenvectors in terms of the basis $|m)_e$.

In the following sections, we shall illustrate how this works for two special
types of unitary representations.

\setcounter{equation}{0}
\section{The Fock representations $W([p,0,\ldots,0])\equiv W(p)$}
\label{sec:Wp}

One interesting class of representations~\cite{Palev3} of $\gl(1|n)$ is that with 
$[m]_{n+1}=[p,0,\ldots,0]$, i.e.\ with highest weight $\Lambda = p \epsilon$.
The representation space $W([p,0,\ldots,0])$ is simply denoted by $W(p)$.
It follows from (U1)-(U2) that $W(p)$ is unitary when either $p>n-1$ (typical case) 
or else $p=0,1,\ldots,n-1$ (atypical of type~$p+1$).
In the notation of~(\ref{m}), the GZ-patterns of $W(p)$ consist of zeros and ones only
(apart from the label~$p$), so it will be
convenient to use a simpler notation for these vectors.
The GZ-basis vectors of $W(p)$ will simply be denoted by $w(\varphi_1,\ldots,\varphi_n)\equiv w(\varphi)$, where
the relation to the GZ-labels is determined by~\cite{KSV}
\begin{equation}
\varphi_i= \sum_{j=1}^i m_{ji} - \sum_{j=1}^{i-1} m_{j,i-1}.
\end{equation}
The constraints (GZ2)-(GZ4) for the GZ-labels lead to: $\varphi_i\in\{0,1\}$ and 
$\sum_{i=1}^n \varphi_i \leq \min(p,n)$.
The representations $W(p)$ and the basis vectors $w(\varphi_1,\ldots,\varphi_n)$
have been constructed by means of Fock space techniques,
and the action of the $\gl(1|n)$ generators is very simple, see~\cite{Palev3}.
The Fock construction gives all vectors in terms of the highest weight vector
$|\Lambda)_e\equiv w(0,\ldots,0)$:
\begin{equation*}
w(\varphi) = w(\varphi_1,\ldots,\varphi_n) =
\frac{e_{10}^{\varphi_1}e_{20}^{\varphi_2}\cdots e_{n0}^{\varphi_n}}{\sqrt{p(p-1)\cdots(p-|\varphi|+1)}} w(0,\ldots,0),
\end{equation*}
where $|\varphi|=\sum \varphi_i$. 
The action of the $e_{ij}$ on such vectors is determined by ($1\leq k\leq n$):
\begin{align}
e_{00} w(\varphi) &= (p-|\varphi|)\ w(\varphi), \label{e00} \\
e_{kk} w(\varphi) &= \varphi_k\ w(\varphi), \label{ekk} \\
e_{k0} w(\varphi) &= (1-\varphi_k) (-1)^{\varphi_1+\cdots+\varphi_{k-1}} \sqrt{p-|\varphi|}\
 w(\varphi_1,\ldots, \varphi_k+1,\ldots, \varphi_n), \label{ek0} \\
e_{0k} w(\varphi) &= \varphi_k (-1)^{\varphi_1+\cdots+\varphi_{k-1}} \sqrt{p-|\varphi|+1}\
 w(\varphi_1,\ldots, \varphi_k-1,\ldots, \varphi_n). \label{e0k}
\end{align}

Now we introduce the second GZ-basis, denoted by $|m)_E$ in the previous paragraphs. 
Analogously to the previous basis, we shall use a simpler notation, namely
$v(\phi) = v(\phi_1,\ldots,\phi_n)$, with each $\phi_i \in\{0,1\}$.  This basis is defined by:
\begin{equation}
\label{v-phi}
v(\phi) = v(\phi_1,\ldots,\phi_n) \equiv
\frac{E_{10}^{\phi_1}E_{20}^{\phi_2}\cdots E_{n0}^{\phi_n}}{\sqrt{p(p-1)\cdots(p-|\phi|+1)}} v(0,\ldots,0),
\end{equation}
where $E_{j0}$ is determined by~(\ref{En0}) and~(\ref{Ej0}), and $v(0,\ldots,0)$ is
the highest weight vector $|\Lambda)_E$ with respect to the $E_{ij}$ basis of $\gl(1|n)$.
This vector is given by~(\ref{E-hw}). However, in the current case there is only one vector of 
the type~(\ref{md}), so $|\Lambda)_E=|\Lambda)_e$, in other words: $v(0,\ldots,0)=w(0,\ldots,0)$.
This implies, in particular, that $v(\phi)$ is a linear combination of vectors $w(\varphi)$
with $|\varphi|=|\phi|$. 

For the typical case, all $N$-values in~(\ref{branching}) are~1, and the decomposition becomes
\begin{equation}
W(p) \rightarrow \bigoplus_{K=0}^{n-1} W([p-K,0]) \times V([\underbrace{1,\ldots,1}_{K},0,\ldots,0]),
\end{equation}
where the $\gl(n-1)$ representation $V([1,\ldots,1,0,\ldots,0])$ has $K$ ones and $n-1-K$ zeros,
with $\dim V([1,\ldots,1,0,\ldots,0]) = \binom{n-1}{K}$. Consequently, $\q_r$ has $2n$ eigenvalues
$\pm x_K = \pm \sqrt{\frac{\hbar\gamma}{\mu n}(p-K)}$, where $0 \leq K \leq n-1$, with multiplicities
$\binom{n-1}{K}$. 
The orthonormal eigenvectors are:
\begin{equation}
\label{psi}
\psi_{r,\pm x_K,\phi} = \frac{1}{\sqrt{2}} v(\phi_1,\ldots,\phi_{n-1},0)
\pm \frac{(-1)^{\phi_1+\cdots+\phi_{n-1}}}{\sqrt{2}} v(\phi_1,\ldots,\phi_{n-1},1),
\end{equation}
where $\phi_1+\cdots+\phi_{n-1}=K$.
It is indeed easy to check, with $\q_r=\sqrt{\frac{\hbar\gamma}{\mu n}}(E_{0n}+E_{n0})$, that
\begin{equation}
\q_r \psi_{r,\pm x_K,\phi} = \pm \sqrt{\frac{\hbar\gamma (p-K)}{\mu n}}\psi_{r,\pm x_K,\phi}.
\end{equation}

Thus we have:
\begin{prop}
In the typical representation $W(p)=W([p,0,\ldots,0])$ ($p>n-1$), 
the operator $\q_r$ has $2n$ distinct eigenvalues 
given by $\pm x_K=\pm \sqrt{\frac{\hbar\gamma}{\mu n}(p-K)}$, 
where $0 \leq K \leq n-1$.
The multiplicity of the eigenvalue $\pm x_K$ is $\binom{n-1}{K}$.  
The eigenvectors of $\q_r$ for the eigenvalue $\pm x_K$ contain, when expanded
in the standard basis $w(\varphi)$, only vectors with 
$|\varphi|=K$ or $|\varphi|=K+1$.  A set of orthonormal eigenvectors
is given by~\eqref{psi}. 
\end{prop}

What happens in the atypical case? Then $p\in\{0,1,\ldots,n-1\}$ and $W(p)$
is atypical of type~$p+1$. 
Now the decomposition~(\ref{branching}) becomes
\begin{equation}
W(p) \rightarrow \bigoplus_{K=0}^{p} W([p-K,0]) \times V([\underbrace{1,\ldots,1}_{K},0,\ldots,0]).
\end{equation}
Consequently, $\q_r$ has $2p$ nonzero eigenvalues
$\pm x_K = \pm \sqrt{\frac{\hbar\gamma}{\mu n}(p-K)}$, where $0 \leq K \leq p-1$, with multiplicities
$\binom{n-1}{K}$; and one zero eigenvalue $x_p=0$ with multiplicity $\binom{n-1}{p}$.
For a nonzero eigenvalue, the orthonormal eigenvectors take the same
form as~\eqref{psi}. For the zero eigenvalue, the orthonormal
eigenvectors are simply all vectors $v(\phi)$ with $|\phi|=p$ and $\phi_n=0$.

Note that the spectrum of $\q_r$ is independent of $r$, i.e.\ independent of the
location of the oscillator in the linear chain of $n$ oscillators.
The eigenvectors, however, do depend on~$r$. 
This is because in~(\ref{v-phi}) the generators $E_{j0}$ do indeed depend on~$r$,
see~(\ref{En0}) and (\ref{Ej0}).

Using (\ref{psi}), (\ref{v-phi}), (\ref{En0}) and (\ref{Ej0}), one can explicitly compute the coefficients
\begin{equation}
\psi_{r,\pm x_{|\phi|},\phi} = 
\sum_{\varphi} C_{r,\pm x_{|\phi|},\phi}^{\varphi} w(\varphi)
\label{expansion}
\end{equation}
for the expansion of the $\q_r$ eigenvectors in terms of the stationary states $w(\varphi)$.
We have already noted that in the right hand side of~(\ref{expansion}), only terms with
$|\varphi|=|\phi|$ or $|\varphi|=|\phi|+1$ can be nonzero. 

When the quantum system is in a fixed eigenstate $w(\varphi)$ of $\hat H$,
then the probability of measuring for $\q_r$ the eigenvalue $\pm x_K$ is given by
\begin{equation}
\label{prob}
P(\varphi,r,\pm x_K) = \sum_{\phi,\, \phi_1+\phi_2+\cdots+\phi_{n-1}=K}
 | C^{\varphi}_{r,\pm x_K,\phi}|^2.
\end{equation}
Without giving details of the computations, we have deduced:
\begin{equation}
P(\varphi,r, x_K) = 
\begin{cases}
\frac{1}{2\gamma} \sum_{j=1}^n (1-\varphi_j) \gamma_j^2
 & \text{when}\ |\varphi| = K \\
\frac{1}{2\gamma} \sum_{j=1}^n \varphi_j \gamma_j^2
 & \text{when}\ |\varphi| = K+1, \\
 0 & \text{otherwise.}
\end{cases}
\label{prob1}
\end{equation}
Since
\[
P(\varphi,r, -x_K) = P(\varphi,r, x_K),
\]
one has
\begin{equation*}
	P(\varphi,r,x_{|\varphi|}) + P(\varphi,r,x_{|\varphi|-1})+ 
	P(\varphi,r,-x_{|\varphi|}) + P(\varphi,r,-x_{|\varphi|-1}) = 1,
\end{equation*}
implying the following: when the quantum system is in the stationary state $w(\varphi)$,
a measurement of $\q_r$ leads to four possible values $\pm x_{|\varphi|}, \pm x_{|\varphi|-1}$,
with probabilities given by~(\ref{prob}) and~(\ref{prob1}).

\setcounter{equation}{0}
\section{The ladder representations $W([1,p-1,0,\ldots,0])\equiv V(p)$}
\label{sec:Vp}

Another interesting class of representations~\cite{KSV} of $\gl(1|n)$ is that with 
$[m]_{n+1}=[1,p-1,0,\ldots,0]$, denoted by $V(p)$.
By~(GZ1), $p$ is a positive integer, and by~(U2) it is a unitary representation
atypical of type~2.
The notation~(\ref{m}) for the GZ-patterns of $V(p)$ have again too many zeros to
be convenient, so the vectors will be denoted in a simpler way.
In this case, one can write the $|m)_e$'s as $w(\theta;s_1,s_2\ldots,s_n)\equiv w(\theta;s)$, where
\begin{equation}
\theta=p-m_{1n},\quad s_1=m_{11},\quad s_k=m_{1k}-m_{1,k-1}\quad (k=2,\ldots, n).
\end{equation}
Thus all vectors of $V(p)$ are described by:
\begin{equation}
w(\theta;s)\equiv w(\theta;s_1,s_2, \ldots,s_n),\qquad
\theta\in\{0,1\},\ s_i\in\{0,1,2,\ldots\},\ \hbox{ and } \theta+s_1+\cdots+s_n= p.
\label{newbasis}
\end{equation}
In this notation the highest weight vector is $|\Lambda)_e=w(1;p-1,0,\ldots,0)$.
The action of the $\gl(1|n)$ generators on the new basis~(\ref{newbasis}) is given 
by ($1\leq k \leq n$):
\begin{align}
& e_{00} w(\theta; s) = \theta\ w(\theta; s),\\
& e_{kk} w(\theta; s) = s_k\ w(\theta; s),\\
& e_{k0} w(\theta; s) =\theta \sqrt{s_k+1}\ w(1-\theta; s_1,\ldots,s_k+1,\ldots,s_n), \\
& e_{0k} w(\theta; s) =(1-\theta)\sqrt{s_k}\ w(1-\theta; s_1,\ldots,s_k-1,\ldots,s_n).
\end{align}
From these one deduces the action of other elements $e_{kl}$.
The ladder representations $V(p)$ and the basis vectors $w(\theta;s_1,s_2\ldots,s_n)$
can also be constructed by means of negative root vectors acting on the
highest weight vector. In particular:
\begin{equation*}
w(\theta;s_1,\ldots,s_n) =  
\frac{ 
e_{n,n-1}^{p-\theta-\sum_{j=1}^{n-1}s_j} 
e_{n-1,n-2}^{p-\theta-\sum_{j=1}^{n-2}s_j} 
\cdots  
e_{32}^{p-\theta-\sum_{j=1}^2 s_j} 
e_{21}^{p-\theta-\sum_{j=1}^1 s_j} 
e_{10}^{1-\theta} }
{
\sqrt{ p^{1-\theta} \prod_{k=1}^{n-1} (p-\theta - \sum_{j=1}^k s_j)! (s_k+1)_{p-\theta - \sum_{j=1}^k s_j}  }
}
w(1;p-1,0,\ldots,0),
\end{equation*}
where $(a)_j=a(a+1)\cdots (a+j-1)$ is the Pochhammer symbol or rising factorial.

Now we also introduce the second GZ-basis $|m)_E$, but in the same
simpler notation, namely
$v(\phi;t) = v(\phi;t_1,\ldots,t_n)$, with $\phi\in\{0,1\}$, $t_i\in\Z_+$ and $\phi+t_1+\cdots+t_n=p$.  
This basis is defined by:
\begin{equation}
\label{Vv-phi}
v(\phi;t_1,\ldots,t_n) =
\frac{ 
E_{n,n-1}^{p-\phi-\sum_{j=1}^{n-1}t_j} 
E_{n-1,n-2}^{p-\phi-\sum_{j=1}^{n-2}t_j} 
\cdots  
E_{32}^{p-\phi-\sum_{j=1}^2 t_j} 
E_{21}^{p-\phi-\sum_{j=1}^1 t_j} 
E_{10}^{1-\phi} }
{
\sqrt{ p^{1-\phi} \prod_{k=1}^{n-1} (p-\phi - \sum_{j=1}^k t_j)! (t_k+1)_{p-\phi - \sum_{j=1}^k t_j}  }
}
v(1;p-1,0,\ldots,0).
\end{equation}
where $E_{j0}$ is determined by~(\ref{En0}) and~(\ref{Ej0}), 
$E_{j+1,j}=\{ E_{j+1,0},E_{0,j}\}$,
and $v(1;p-1,0,\ldots,0)$ is
the highest weight vector $|\Lambda)_E$ with respect to the $E_{ij}$ basis of $\gl(1|n)$.
In general, this vector is given by~(\ref{E-hw}), and here this becomes:
\begin{align}
v(1;p-1,0,\ldots,0)& =\frac{1}{(\gamma_1^2+\gamma_2^2)^{(p-1)/2}}
\sum_{u=0}^{p-1} (-1)^{u} e^{-2\pi i r u/n} \sqrt{\binom{p-1}{u}}\nn\\
&\qquad \times \gamma_1^{p-1-u}\gamma_2^{u}\; w(1;u,p-1-u,0,\ldots,0).
\label{V-LambdaE}
\end{align}

The decomposition~(\ref{branching}) reads,
\begin{equation}
V(p) \rightarrow W([0,0])\times V([p,0,\ldots,0]) \oplus \bigoplus_{K=0}^{p-1} W([1,p-1-K]) \times V([K,0,\ldots,0]),
\end{equation}
where the $\gl(n-1)$ representation has $\dim V([K,0,\ldots,0]) = \binom{n-2+K}{n-2}$. 
So $\q_r$ has $2p+1$ eigenvalues in all, namely
$\pm x_K = \pm \sqrt{\frac{\hbar\gamma}{\mu n}(p-K)}$, where $0 \leq K \leq p-1$, with multiplicities
$\binom{n-2+K}{n-2}$, and $x_p=0$ with multiplicity $\binom{n-2+p}{n-2}$. 
The orthonormal eigenvectors for $\pm x_K\ne 0$ are:
\begin{equation}
\label{Vpsi}
\psi_{r,\pm x_K,t} = \frac{1}{\sqrt{2}} v(1;t_1,\ldots,t_{n-1},p-1-K)
+\frac{1}{\sqrt{2}} v(0;t_1,\ldots,t_{n-1},p-K),
\end{equation}
where $t_1+\cdots+t_{n-1}=K$. 
For the eigenvalue~0, the eigenvectors read
\begin{equation}
\label{Vpsi0}
\psi_{r,0,t}= v(0; t_1,\ldots,t_{n-1},0),\qquad t_1+\cdots+t_{n-1}=p.
\end{equation}
In other words:
\begin{prop}
In the representation $V(p)=W([1,p-1,0,\ldots,0])$, 
the operator $\q_r$ has $2p+1$ distinct eigenvalues 
given by $\pm x_K=\pm \sqrt{\frac{\hbar\gamma}{\mu n}(p-K)}$, 
where $0 \leq K \leq p$.
The multiplicity of the eigenvalue $\pm x_K$ is $\binom{n-2+K}{K}$.  
A set of orthonormal eigenvectors is given by~\eqref{Vpsi} and~\eqref{Vpsi0}. 
\end{prop}

\setcounter{equation}{0}
\section{Conclusions}
\label{sec:conclusion}

In this paper we managed to determine the eigenvalues of an arbitrary self-adjoint odd element~\eqref{arbitrary}
of the Lie superalgebra $\gl(1|n)$ in a unitary representation $W=W([m]_{n+1})$.
Furthermore, we gave a construction of a set of orthonormal eigenvectors of this element in $W$,
using the GZ-basis vectors.
The problem is of importance in the study of physical properties of the $\gl(1|n)$ Wigner quantum system solution
for a model consisting of a linear chain of $n$ harmonic oscillators coupled by springs, with periodic boundary
conditions.
In such a description, the position and momentum operator $\q_r$ and $\p_r$ of the $r$th oscillator are such
odd elements, see~(\ref{q-e})-(\ref{p-e}).
We have concentrated on the operator $\q_r$. Note, by~(\ref{p-e}), that the analysis of $\p_r$ is very similar:
one should replace all constants $\gamma_j$ by $\sqrt{\beta_j}$, leading to the analogue of~(\ref{q-E}):
\begin{equation}
\p_r = i \sqrt{\frac{\mu\hbar\beta}{n}} (E_{0n}-E_{n0}), \qquad (\beta=\beta_1+\cdots+\beta_n).
\end{equation}
Then the counterpart of~(\ref{qr-vw}) is
\begin{equation}
\p_r  \frac{v\pm i w}{\sqrt{2}} = \mp \sqrt{\frac{\mu \hbar\beta}{n}}\sqrt{a+b}\ \frac{v\pm i w}{\sqrt{2}}.
\label{pr-vw}
\end{equation}
So, up to an overall factor, the spectrum of $\p_r$ is the same as that of $\q_r$. The eigenvectors, however,
are different, but can be found by a similar construction.

As an application, we have in mind the description of some geometric aspects of the $\gl(1|n)$ solution
of the quantum system described. These aspects depend on the representation considered.
For the simple class of Fock representations $W(p)$, some properties were already described in~\cite{LSV}.
Clearly, the ladder representations $V(p)$ have a much richer structure.
It would be interesting to study such properties for these representations. In particular, we have in mind:
position probability distributions for the stationary states $w(\theta;s)$; position probabilities for the
other oscillators when one oscillator is in an eigenstate with fixed eigenvalue; average position of the
other oscillators when one oscillator is in a fixed position, etc.
For all these aspects, one needs the explicit expansion of the orthonormal $\q_r$ eigenvectors
in terms of the basis of stationary states $w(\theta;s)$, as determined in this paper in Section~\ref{sec:Vp}.

We want to point out that the analysis presented here will be useful not only for 
the quantum system described here in Section~\ref{sec:problem}, but also for the study
of related models. For example, a quantum system consisting of 
a linear chain of harmonic oscillators coupled by springs, but with non-periodic boundary
conditions (i.e.\ with fixed end points) also allows a $\gl(1|n)$ Wigner quantum system solution.
The techniques developed here should be useful in the study of such alternative systems.

\section*{Acknowledgments}
NIS was supported by a project from the Fund for Scientific Research -- Flanders (Belgium).

\appendix 
\section{Appendix}
\renewcommand{\theequation}{\Alph{section}.\arabic{equation}}
\setcounter{equation}{0}

The explicit action of a set of $\gl(1|n)$ generators 
on the basis vectors~(\ref{m}) was given in~\cite[Eq.~(2.13)-(2.18)]{KSV}. 
For the readability of this paper, we repeat this here.
Denote by $|m)_{\pm ij}$ the pattern obtained from $|m)$ by the replacement
$m_{ij} \rightarrow m_{ij}\pm 1$. Then the action is given by:
\begin{align}
e_{00}|m)&=\left(m_{0,n+1}-\sum_{j=1}^n \theta_{j}\right)|m); \label{e_00}\\
e_{kk}|m)&=\left(\sum_{j=1}^k m_{jk}-\sum_{j=1}^{k-1} m_{j,k-1}\right)|m), 
\quad (1\leq k\leq n); \label{e_kk}\\
e_{k-1,k}|m)&=\sum_{j=1}^{k-1} \left(-
\frac{\prod_{i=1}^{k} (l_{ik}-l_{j,k-1})
\prod_{i=1}^{k-2} (l_{i,k-2}-l_{j,k-1}-1)}{\prod_{i\neq j=1}^{k-1} (l_{i,k-1}-l_{j,k-1})
(l_{i,k-1}-l_{j,k-1}-1) } \right)^{1/2}|m)_{+j,k-1},\quad (2\leq k\leq n); \nn\\[-3mm]
& \label{ek}\\
e_{k,k-1}|m)&=\sum_{j=1}^{k-1} \left(-
\frac{\prod_{i=1}^{k} (l_{ik}-l_{j,k-1}+1)
\prod_{i=1}^{k-2} (l_{i,k-2}-l_{j,k-1})}{\prod_{i\neq j=1}^{k-1} (l_{i,k-1}-l_{j,k-1})
(l_{i,k-1}-l_{j,k-1}+1) } \right)^{1/2}|m)_{-j,k-1},\quad (2\leq k\leq n); \nn\\[-3mm]
& \label{fk}\\
e_{0n}|m)&=\sum_{i=1}^n \theta_{i}
(-1)^{\theta_{1}+ \ldots +\theta_{i-1} }(l_{i,n+1}+l_{0,n+1}+1)^{1/2}
\left( \frac{\prod_{k=1}^{n-1}  (l_{k,n-1}-l_{i,n+1}-1 )}{\prod_{k\neq i=1}^n (  l_{k,n+1}-l_{i,n+1})}
\right)^{1/2} |m)_{-in}; \nn \\
& \label{en}\\
e_{n0}|m)&=\sum_{i=1}^n (1-\theta_{i})
(-1)^{\theta_{1}+ \ldots +\theta_{i-1} }(l_{i,n+1}+l_{0,n+1}+1)^{1/2}\nn\\
 & \qquad \times \left(
\frac{\prod_{k=1}^{n-1} ( l_{k,n-1}-l_{i,n+1} -1)}{\prod_{k\neq i=1}^n (  l_{k,n+1}-l_{i,n+1} )}
\right)^{1/2} |m)_{+in}.\label{fn}
\end{align}
In all these formulas $l_{ij}=m_{ij}-i$.

It is also useful to know the explicit action of all the odd elements
$e_{0j}$ and $e_{j0}$ of $\gl(1|n)$. This was found in~\cite[Eq.~(2.25)-(2.26)]{KSV}:
\begin{align}
e_{0j}|m)&=\sum_{i_n=1}^n\sum_{i_{n-1}=1}^{n-1}\ldots \sum_{i_j=1}^j
\theta_{i_n}(-1)^{\theta_1+\cdots +\theta_{i_n-1}}(l_{i_n,n+1}+l_{0,n+1}+1)^{1/2}\nn\\
&\times
\prod_{r=j+1}^nS(i_r,i_{r-1})
\left( \frac{\prod_{k\neq i_{r-1}=1}^{r-1}  
(l_{k,r-1}-l_{i_r,r} )\prod_{k\neq i_r=1}^r(l_{kr}-l_{i_{r-1},r-1}+1)
 }{ \prod_{k\neq i_r=1}^r (l_{kr}-l_{i_r,r})\prod_{k\neq i_{r-1}=1}^{r-1}(l_{k,r-1}
 -l_{i_{r-1},r-1}+1)}
\right)^{1/2} \label{e0j}\\
&\times 
\left({\prod_{k\neq i_n=1}^n } \frac{(l_{kn}-l_{i_n,n} )}{(  l_{k,n+1}-l_{i_n,n+1})}
\right)^{1/2}
\left( \frac{\prod_{k=1}^{j-1} (l_{k,j-1}-l_{i_j,j} )}{\prod_{k\neq i_j=1}^j(  l_{kj}-l_{i_j,j})}
\right)^{1/2}
|m)_{-i_n,n;-i_{n-1},n-1;\ldots ;-i_j,j}\nn
\end{align}

\begin{align}
e_{j0}|m)&=\sum_{i_n=1}^n\sum_{i_{n-1}=1}^{n-1}\ldots \sum_{i_j=1}^{j}
(1-\theta_{i_n})(-1)^{\theta_1+\cdots +\theta_{i_n-1}}(l_{i_n,n+1}+l_{0,n+1}+1)^{1/2}\nn\\
&\times
\prod_{r=j+1}^nS(i_r,i_{r-1})
\left( \frac{\prod_{k\neq i_{r-1}=1}^{r-1}  
(l_{k,r-1}-l_{i_r,r}-1 )\prod_{k\neq i_r=1}^r(l_{kr}-l_{i_{r-1},r-1})
 }{  \prod_{k\neq i_r=1}^r (l_{kr}-l_{i_r,r})\prod_{k\neq i_{r-1}=1}^{r-1}(l_{k,r-1}
 -l_{i_{r-1},r-1}-1)}
\right)^{1/2} \label{ej0}\\
&\times 
\left({\prod_{k\neq i_n=1}^n } \frac{(l_{kn}-l_{i_n,n} )}{(  l_{k,n+1}-l_{i_n,n+1})}
\right)^{1/2}
\left( \frac{\prod_{k=1}^{j-1} (l_{k,j-1}-l_{i_j,j}-1 )}{\prod_{k\neq i_j=1}^j(  l_{kj}-l_{i_j,j})}
\right)^{1/2}
|m)_{+i_n,n;+i_{n-1},n-1;\ldots ;+i_j,j},\nn
\end{align}
where $j=1,\ldots ,n$, each symbol $\pm i_k,k$ attached as a subscript to $|m)$ indicates a
replacement $m_{i_k,k}\rightarrow m_{i_k,k}\pm 1$, and 
\begin{equation}
S(k,l) = \left\{ \begin{array}{lll}
 {\;\;1} & \hbox{for} & k\leq l  \\ 
 {-1} & \hbox{for} & k>l .
 \end{array}\right.
\end{equation}


\end{document}